\documentclass[10pt,conference]{IEEEtran}
\IEEEoverridecommandlockouts

\usepackage[T1]{fontenc}
\usepackage[utf8]{inputenc}
\usepackage{microtype}

\usepackage{amsmath,amssymb,amsfonts,mathtools}
\usepackage{bm}
\usepackage{amsthm}

\usepackage{graphicx}
\graphicspath{{gfx/}{./gfx/}}

\usepackage[caption=false,font=footnotesize]{subfig}
\usepackage{booktabs}
\usepackage{tabularx}
\usepackage{enumitem}
\usepackage{balance}

\usepackage{tikz}
\usepackage[percent]{overpic}
\usepackage{quantikz}
\usepackage{cite}
\ifdefined\CPSBuild
  \PassOptionsToPackage{draft,bookmarks=false}{hyperref}
\fi
\definecolor{gateH}{RGB}{214, 234, 248}
\definecolor{gateHborder}{RGB}{40, 116, 166}
\definecolor{gateU}{RGB}{250, 219, 216}
\definecolor{gateUborder}{RGB}{176, 58, 46}
\definecolor{edgeNN}{RGB}{33, 97, 140}
\definecolor{edgeNNN}{RGB}{176, 58, 46}
\usepackage{hyperref}
\usepackage{cleveref}
\usepackage{orcidlink}

\def\BibTeX{{\rm B\kern-.05em{\sc i\kern-.025em b}\kern-.08em
T\kern-.1667em\lower.7ex\hbox{E}\kern-.125emX}}

\newcommand{\figplaceholder}[1]{%
  \fbox{\parbox[c][0.18\textheight][c]{0.95\linewidth}{%
    \centering\footnotesize
    Figure file not found:\\[2pt]\texttt{\detokenize{#1}}%
  }}%
}
\newcommand{\maybeincludegraphics}[2][]{%
  \IfFileExists{#2}{\includegraphics[#1]{#2}}{%
    \IfFileExists{gfx/#2}{\includegraphics[#1]{gfx/#2}}{%
      \IfFileExists{./gfx/#2}{\includegraphics[#1]{./gfx/#2}}{%
        \figplaceholder{#2}%
      }%
    }%
  }%
}

\newcommand{\cS}{\mathcal{S}}

\newcommand{\cD}{\mathcal{D}}
\newcommand{\cH}{\mathcal{H}}

\newcommand{\cE}{\mathcal{E}}
\newcommand{\cO}{\mathcal{O}}
\newcommand{\cU}{\mathcal{U}}

\newcommand{\ind}{\mathbf{1}}

\theoremstyle{definition}

\theoremstyle{remark}

\hypersetup{
  pdftitle={Parity Supervision as a Driver of Generalization in Quantum Generative Modeling},
  pdfauthor={Markus Baumann, Daniel Hein, Steffen Udluft, Tobias Rohe, Claudia Linnhoff-Popien, Jonas Stein},
  pdfsubject={Parity supervision and generalization in IQP quantum circuit Born machines},
  pdfkeywords={instantaneous quantum polynomial-time circuits, parity supervision, quantum circuit Born machines, quantum generative modeling, quantum machine learning}
}
\ifdefined\CPSBuild
\else
  \hypersetup{
    colorlinks=true,
    linkcolor={green!80!black},
    citecolor={green!80!black},
    urlcolor={blue!70!black}
  }
\fi

\Crefname{section}{Sec.}{Secs.}
\Crefname{theorem}{Thm.}{Thms.}
\Crefname{lemma}{Lem.}{Lems.}
\Crefname{corollary}{Cor.}{Cors.}
\Crefname{definition}{Def.}{Defs.}
\Crefname{remark}{Rem.}{Rems.}
\Crefname{figure}{Fig.}{Figs.}
\Crefname{equation}{Eq.}{Eqs.}
\Crefname{table}{Table}{Tables}

\DeclareRobustCommand{\IEEEauthorrefmark}[1]{\smash{\textsuperscript{\footnotesize #1}}}

\begin{document}

\ifdefined\CPSBuild\else
\AddToHookNext{shipout/foreground}{%
  \begin{tikzpicture}[remember picture,overlay]
    \node[anchor=south,align=left,text width=7.15in,font=\fontsize{5.5}{6.3}\selectfont]
      at ([yshift=0.18in]current page.south) {\textcopyright~2026 IEEE. Personal use of this material is permitted. Permission from IEEE must be obtained for all other uses, in any current or future media,\\
      including reprinting/republishing this material for advertising or promotional purposes, creating new collective works, for resale or redistribution to servers or lists,\\
      or reuse of any copyrighted component of this work in other works.};
  \end{tikzpicture}%
}
\fi

\title{Parity Supervision as a Driver of Generalization in Quantum Generative Modeling
}

\author{%
\IEEEauthorblockN{%
Markus Baumann\IEEEauthorrefmark{1}\orcidlink{0009-0007-3575-1006}\thanks{Corresponding author: \href{mailto:markus.baumann@campus.lmu.de}{markus.baumann@campus.lmu.de}.}, Daniel Hein\IEEEauthorrefmark{2}\orcidlink{0000-0002-8375-1592},
Steffen Udluft\IEEEauthorrefmark{2}\orcidlink{0000-0002-5767-2591}\\
Tobias Rohe\IEEEauthorrefmark{1}\orcidlink{0009-0003-3283-0586}, Claudia Linnhoff-Popien\IEEEauthorrefmark{1}\orcidlink{0000-0001-6284-9286},
and Jonas Stein\IEEEauthorrefmark{1}\orcidlink{0000-0001-5727-9151}}
\IEEEauthorblockA{\IEEEauthorrefmark{1}\textit{QAR-Lab, Department of Computer Science, LMU Munich, Munich, Germany}}
\IEEEauthorblockA{\IEEEauthorrefmark{2}\textit{Siemens AG, Munich, Germany}}
}

\maketitle

\bstctlcite{BSTcontrol}

\begin{abstract}
Generative models learn probability distributions in order to produce new samples beyond a finite training set. Their usefulness therefore depends on assigning probability to valid but previously unseen states. In a controlled benchmark, we test whether parity-based training provides an inductive bias for this kind of generalization in instantaneous quantum polynomial-time (IQP) circuit Born machines. We compare the same IQP circuit trained with parity supervision and coordinate-wise mean-squared error (MSE), together with classical controls. Parity supervision improves exact distributional fit and recovery of unseen high-value states over IQP-MSE. A circuit-free spectral reconstruction shows that the matched parity moments already transfer evidence from observed samples to structurally compatible unseen states, while the IQP circuit further refines this structure. These results identify parity supervision as both a tractable training signal and a generalization mechanism when the target distribution, training objective, and circuit architecture are spectrally aligned.
\end{abstract}

\begin{IEEEkeywords}
Instantaneous quantum polynomial-time (IQP) circuits, parity supervision, quantum circuit Born machines, quantum generative modeling, quantum machine learning.
\end{IEEEkeywords}

\section{Introduction}
\label{sec:introduction}

Generative models learn to create new samples that reproduce the structure of observed data, supporting tasks such as scientific design, simulation, and combinatorial search. Quantum generative modeling asks whether quantum states and measurement can provide useful model classes or training mechanisms for such discrete distributions.

\begin{figure*}[t]
\centering
\maybeincludegraphics[width=0.9\linewidth]{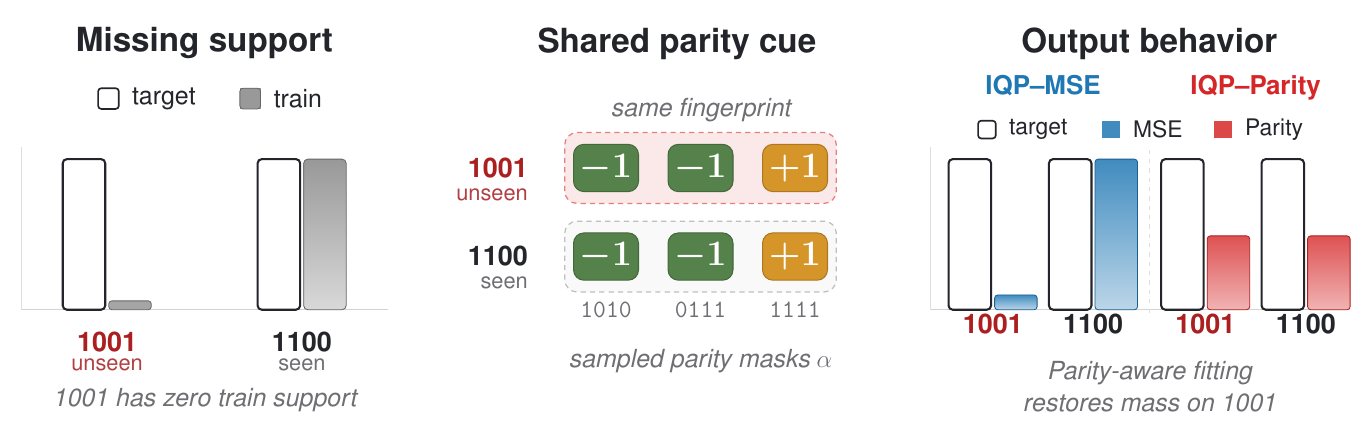}
\caption{How parity supervision can transfer evidence to unseen states. An unobserved valid state (\texttt{1001}) shares the same parity fingerprint as an observed state (\texttt{1100}) across sampled masks. Matching these collective statistics can inform the unseen state through shared parity evidence, whereas unsmoothed coordinate-wise fitting gives the unobserved state no direct positive sample-frequency target.}
\label{fig:intuition-parity-holdout}
\end{figure*}

A generative model trained on finite data is useful only if it assigns probability mass to states that are consistent with the target distribution but absent from the training sample~\cite{SanchezLengeling2018Inverse,Brown2019GuacaMol,Polykovskiy2020MOSES,Bengio2021MachineLearningCO}. The central question is therefore not only whether the model fits the observed histogram, but whether it generalizes to unseen valid states. In realistic discrete domains this question is difficult to answer cleanly: the full target distribution is unknown, and common proxy scores conflate fidelity, diversity, and downstream utility~\cite{Theis2015EvalGM,Sajjadi2018PR,Kynkaanniemi2019PR,Naeem2020PRDC,Gili2022GeneralizationMetrics,HibatAllah2024PQA}.

Quantum circuit Born machines (QCBMs)~\cite{Liu2018Born,Benedetti2019GenerativeBenchmark,Zhu2019TrainingHybridQCircuits} provide a natural quantum model class for discrete distributions: normalization, discrete support, and sampling are built into the Born rule. Within this family, Instantaneous Quantum Polynomial-time (IQP) circuits are particularly relevant for parity-based training because their commuting diagonal structure makes subset-parity correlations native observables. At the same time, prior work often changes the circuit family, loss, and benchmark simultaneously~\cite{Gili2023QCBMGeneralize,Rudolph2024TrainabilityBarriers}, which makes it hard to identify whether a performance gap comes from the architecture, the objective, or the alignment between them.

This paper studies that attribution problem directly. We ask whether parity/Walsh supervision acts as a spectral inductive bias for IQP Born machines, and whether the IQP parameterization contributes beyond the information already contained in the parity moments. The benchmark is intentionally controlled, enumerable, and parity-structured. It is not intended to model a realistic application dataset. Instead, it deliberately trades external validity for identifiability: exact target access makes exact evaluation of the forward Kullback--Leibler (KL) divergence, unseen-state recovery, and within-IQP loss attribution together with diagnostic cross-model comparisons possible.

The core training signal consists of \emph{parity moments}. For a subset of bit positions, a parity moment measures whether the selected bits tend to have even or odd parity. The full set of such moments is the Walsh--Hadamard spectrum of the distribution~\cite{ODonnell2014,Terras1999}. Matching a sampled band of moments is therefore a spectral objective. The contrast is not that mean-squared error (MSE) is an artificially weak baseline. Rather, for any unsmoothed coordinate-wise empirical objective whose per-state targets are sample frequencies, an unobserved state has no positive per-state sample-frequency target in the finite training set. A parity moment has a different information geometry: it couples all states simultaneously through a shared Walsh character. As illustrated in \Cref{fig:intuition-parity-holdout}, observed and unobserved states can share statistical evidence whenever their parity fingerprints agree on the sampled masks.

We instantiate this idea in two complementary controlled contrasts. First, we hold the IQP circuit fixed and replace only the loss, comparing parity supervision with coordinate-wise MSE. Second, we give the same parity information to classical parity-aware controls and compare against additional classical baselines. The within-IQP comparison isolates the supervision effect. The cross-model comparison probes how model class and objective shape the result because the same classical architecture is not trained with both objectives. Within the tested setting, parity supervision explains much of the unseen-state recovery effect, while the IQP parameterization refines the resulting spectral scaffold into the strongest distributional fit among the tested models.

This scope is chosen to enable clean attribution rather than to claim practical quantum advantage. We also do not claim that the tested baselines exhaust all possible classical parity-aware models. The contribution is a controlled mechanism study: by choosing a small, exactly enumerable benchmark, we can measure quantities that would otherwise be replaced by proxies and confounded by model class, objective, optimization, and data availability.

Our contributions are fivefold.
\begin{enumerate}
  \item \textbf{Parity supervision as a spectral inductive bias.}
    We show that matching Walsh--Hadamard parity moments provides an
    inductive bias for finite-sample generalization by constraining
    observed and unseen states jointly.

  \item \textbf{Controlled objective comparison and architecture diagnostic.}
    By swapping only the loss inside the same IQP circuit, we isolate
    the supervision effect within IQP. Matched parity-aware classical
    controls then probe how model class shapes the result.

  \item \textbf{Recovery of unseen high-value states.}
    The improved fit translates into finite-budget discovery: the
    parity-trained IQP model samples more distinct high-scoring unseen
    states than the tested controls.

  \item \textbf{Mechanism via spectral completion.}
    A parameter-free partial Walsh reconstruction already transfers
    evidence to structurally consistent unseen states, showing that much
    of the effect is encoded in the training signal.

  \item \textbf{Hardware feasibility check.}
    On IBM hardware, parity supervision retains its recovery advantage
    over the IQP-MSE control, serving as a feasibility check rather than
    an advantage claim.
\end{enumerate}

\section{Background}
\label{sec:background}
\subsection{IQP Born Machines}
\label{sec:background:iqp}

Quantum circuit Born machines (QCBMs) represent discrete distributions
directly through Born-rule probabilities~\cite{Liu2018Born,%
Benedetti2019GenerativeBenchmark,Zhu2019TrainingHybridQCircuits}.
For a parameterized unitary $U(\bm\theta)$ on $n$ qubits, the model
distribution is

\begin{equation}
  q_{\theta}(x)=\bigl|\langle x|U(\bm\theta)|0^n\rangle\bigr|^2,
  \qquad \text{for } x\in\{0,1\}^n.
  \label{eq:qcbm-definition}
\end{equation}

Thus, normalization, discrete support, and efficient sampling on
the Boolean cube are built into the model by construction.

\begin{figure}[t]
    \centering
    \input{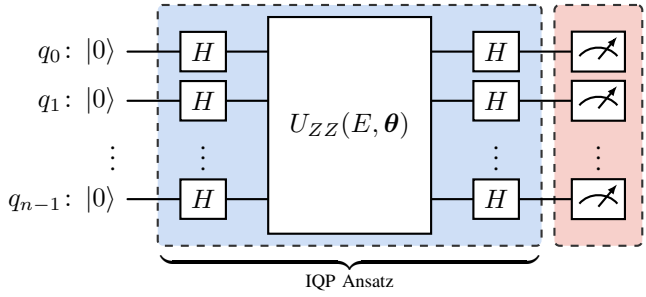}
    \caption{IQP ansatz used throughout the paper. Hadamard layers
    surround a diagonal commuting block $U_{ZZ}(E,\boldsymbol{\theta})$
    with nearest-neighbor (NN) and next-nearest-neighbor (NNN) edge
    sets $E = E_{\mathrm{NN}} \cup E_{\mathrm{NNN}}$, where
    $E_{\mathrm{NN}}=\{(i,i+1 \bmod n)\}$ and
    $E_{\mathrm{NNN}}=\{(i,i+2 \bmod n)\}$ on a ring. Each active $ZZ$
    coupling has its own angle $\theta_{ij}$. Measurement in the
    computational basis yields the Born distribution
    $q_{\boldsymbol{\theta}}(x)$.}
    \label{fig:iqp-circuit}
  \end{figure}

QCBMs form a broad family, and the choice of ansatz determines which
statistics of $q_\theta$ are natively accessible and which training
objectives are tractable. We restrict attention to the
IQP subfamily because its
commuting diagonal structure makes subset-parity correlations among
output bits the \emph{native} observables of the model. The concrete
IQP architecture used throughout this work is shown in
\Cref{fig:iqp-circuit}. It takes the form
\begin{equation}
  U(\bm\theta)=H^{\otimes n}\,D(\bm\theta)\,H^{\otimes n},
  \label{eq:iqp-form}
\end{equation}
with $D(\bm\theta)$ diagonal in the computational basis.

For our purposes, it is important to distinguish two consequences of
this structure. First, the IQP form makes parity moments a natural
training signal: subset-parity correlations are native observables of
the circuit and the corresponding moments entering the loss are
classically tractable. This is precisely why parity supervision is an
attractive objective for IQP Born machines. Second, exact evaluation of
the full $2^n$-dimensional output distribution remains exponential in
$n$. For the small system sizes considered here, we nevertheless exploit
this exact computation---implemented by applying $D(\bm\theta)$
entry-wise and performing two Hadamard transforms---solely for
evaluation in \Cref{sec:results}, where it enables exact
forward-KL computation.

\subsection{Parity Moments and the Walsh Basis}
\label{sec:background:walsh}

For a mask $\alpha\in\{0,1\}^n$, the associated \emph{Walsh character},
i.e.\ the subset-parity function indexed by $\alpha$, is
\begin{equation}
  \phi_{\alpha}(x)=(-1)^{\alpha\cdot x},
  \label{eq:walsh-character}
\end{equation}
where $\alpha\cdot x \equiv \sum_{i=1}^{n}\alpha_i\,x_i \pmod{2}$.
The corresponding \emph{parity moment} of a distribution $r$ on
$\{0,1\}^n$ is
\begin{equation}
  \widehat{r}(\alpha)
  \;=\;\mathbb{E}_{x\sim r}\!\left[\phi_{\alpha}(x)\right]
  \;=\;\sum_{x\in\{0,1\}^n} r(x)\,\phi_{\alpha}(x).
  \label{eq:parity-moment}
\end{equation}

Concretely, the mask $\alpha=0110$ selects bits~2 and~3, so
$\phi_{0110}(x)=(-1)^{x_2+x_3}$ returns $+1$ when the two bits agree and
$-1$ when they disagree; $\widehat{r}(0110)$ thus measures the net
tendency of~$r$ toward agreement or disagreement on this subset. Higher-weight masks probe joint parities over larger subsets, so that a single
expectation value couples many states at once---the structural property
behind the holdout-recovery mechanism shown in
\Cref{fig:intuition-parity-holdout}.
The $2^n$ Walsh characters form an orthonormal basis of the real-valued
functions on $\{0,1\}^n$. Equivalently, the map
$r \mapsto \widehat r$ is invertible, so the full set of parity moments
$\widehat r(\alpha)$ uniquely determines $r$.

The full collection of parity moments $\widehat r(\alpha)$ plays the
role of a Fourier spectrum for distributions on the Boolean cube
$\{0,1\}^n$. More precisely, it is the Walsh--Hadamard spectrum of
$r$---the binary-domain analogue of a Fourier decomposition
\cite{ODonnell2014,Terras1999}. Each coefficient $\widehat r(\alpha)$
measures how strongly the distribution aligns with the subset-parity
pattern indexed by $\alpha$: low-weight masks capture coarse few-bit
correlations, while higher-weight masks capture finer many-bit
structure. Because the Walsh characters form an orthonormal basis, the
full set of coefficients $\widehat r(\alpha)$ uniquely determines $r$,
with inverse reconstruction
\begin{equation}
  r(x)=2^{-n}\sum_{\alpha\in\{0,1\}^n}\widehat r(\alpha)\,\phi_\alpha(x).
\end{equation}

Each
moment reduces to the expectation value of a Pauli-$Z$ string on
quantum hardware and can be estimated to precision~$\epsilon$ from
$O(1/\epsilon^2)$ samples independently of~$n$, a property we exploit
in the hardware experiments of \Cref{sec:results}.

In practice we do not match the full spectrum but a sampled band
$\Omega$ of $K$ nonzero masks. This choice is deliberate: rather than
engineering a task-specific set of masks from domain knowledge, we use
a generic sampled band in order to isolate whether parity supervision
itself already induces generalization.

To construct a parity fingerprint, choose $K$ masks that identify bit
subsets. For every sample and mask, compute $\phi_\alpha(x)$, which is
$+1$ for even parity and $-1$ for odd parity. The $K$ sample averages
form the dataset fingerprint, while the $K$ values for one state form
its fingerprint. We sample masks from $\pi_\sigma$. In applications,
masks could instead represent known constraints or stable correlations.
Otherwise, candidate bands can be compared using training-only or
validation criteria. We do not establish an optimal selection rule.

The key structural property is
that every state $x\in\{0,1\}^n$ enters every matched moment through
$\phi_\alpha(x)$, so even a partial band constrains observed and
unobserved states simultaneously. This coupling is the mechanism that
drives the training effect analyzed in \Cref{sec:methodology:signals}.

\section{Related Work}
\label{sec:related-work}

Three lines of prior work are relevant to our question, but none
directly addresses it.

\paragraph{Parity-based training for QCBMs}
Quantum circuit Born machines (QCBMs), including IQP-based generators,
are established tools for discrete quantum generative
modeling~\cite{Liu2018Born,Benedetti2019GenerativeBenchmark,Zhu2019TrainingHybridQCircuits}.
Early work trained these models with objectives based on maximum mean
discrepancy (MMD)~\cite{Gretton2012MMD}, which treat subset-parity statistics
implicitly through a kernel expansion. More recent IQP-specific
proposals made the parity structure explicit, exploiting the fact
that MMD can be rewritten as a mixture over subset-parity statistics
that IQP circuits evaluate efficiently on classical
hardware~\cite{Rudolph2024TrainabilityBarriers,RecioArmengol2025IQPopt,VanDenNest2011Probabilistic,RecioArmengol2025TrainOnClassical}.
These works motivate parity-based losses on grounds of
\emph{classical simulability} and typically vary model class and
training objective together, so existing results cannot separate
whether observed behavior stems from the generator, the supervision
signal, or their interaction. To our knowledge, no prior work has
established whether parity supervision itself improves generalization
over alternative training signals inside a fixed
architecture.

\paragraph{Spectral perspectives on learning and generalization}
A growing body of work frames generalization through the spectral
decomposition of models and training signals. Classical results show
that kernel methods and wide neural networks exhibit a
\emph{spectral bias} toward low-frequency components of the
target~\cite{Rahaman2019SpectralBias,Xu2019FreqPrinciple,Basri2020FreqBias},
with generalization governed by the alignment between the model's
eigenbasis and the target's spectral
decomposition~\cite{Canatar2021SpectralBias,Bordelon2020Spectrum}.
In the quantum setting, parameterized circuits have been
characterized as truncated Fourier series whose accessible spectrum
is determined by the data
encoding~\cite{SchuldSwekeMeyer2021Fourier,Wiedmann2024FourierVQC,SpectralBiasVQML2025}.
This literature has so far focused on supervised regression with
continuous inputs. Our results can be read as a discrete,
generative-modeling instance of the same alignment principle: the
parity objective operates directly on the Walsh eigenbasis that IQP
circuits natively produce, and our controlled comparisons examine how
performance changes when the loss or model class is varied.

\paragraph{Generalization benchmarks for generative models}
A complementary line of work evaluates generative models through
finite-budget discovery
counts~\cite{Gili2022GeneralizationMetrics,HibatAllah2024PQA} or
task-specific proxy scores in molecular design and combinatorial
search~\cite{Brown2019GuacaMol,Polykovskiy2020MOSES}. Because the
target distribution is unavailable, such proxies conflate the
effects of architecture, training signal, and optimization, and
therefore cannot attribute an observed generalization gap to any
one of them. Our controlled comparisons instead require a benchmark whose
target is known exactly, so that forward KL and unseen-state
coverage become direct observables---trading scale for attribution.

\medskip
This paper addresses the gap left open by these lines. Working on a
benchmark whose target distribution is known exactly, we ask two
questions in sequence: does parity supervision improve generalization in this setting,
and if so, how much of the effect stems from the objective, the architecture,
or their interaction? The design of \Cref{sec:setup:models} combines a
within-IQP loss swap, which isolates the loss effect, with
matched-information classical controls that provide diagnostic evidence
about model-class effects.

\section{Methodology}
\label{sec:methodology}

\Cref{tab:notation} summarizes the core notation used throughout
the benchmark definition and the diagnostic quantities introduced in
this section.

\subsection{Training Signals}
\label{sec:methodology:signals}

A central empirical question of this paper is whether the supervision
signal contributes a distinct generalization effect beyond architecture
alone. To isolate this effect, we compare two objectives that can be
swapped within the same circuit while holding architecture, training
data, and optimization budget fixed: a parity-based objective and a
coordinate-wise objective that fits individual state probabilities.

The parity objective does not ask the model to reproduce each
probability $q_\theta(x)$ individually, but to match a set of summary
statistics that each couple all states at once. These summaries are the
parity moments $\widehat{r}(\alpha)$ of \Cref{sec:background:walsh}. Since we
cannot match all $2^n - 1$ nontrivial moments, we sample a band of $K$
masks
$\Omega = \{\alpha^{(k)}\}_{k=1}^{K} \subset \{0,1\}^n \setminus \{0^n\}$
from a product-Bernoulli distribution $\pi_\sigma$, where each
coordinate is drawn independently as
$\alpha^{(k)}_j \sim \mathrm{Bernoulli}(p_\sigma)$ with
$p_\sigma = \tfrac{1}{2}\bigl(1 - e^{-1/(2\sigma^2)}\bigr)$.
The parameter $\sigma$ controls the mask-weight distribution: smaller
$\sigma$ yields larger $p_\sigma$ and therefore heavier masks on
average, whereas larger $\sigma$ biases $\Omega$ toward lower-weight
masks. The
parity-supervised model then minimizes the mean squared mismatch
between its own moments and those of the training data,
\begin{equation}
  \mathcal{L}_{\mathrm{parity}}(\boldsymbol{\theta})
    = \frac{1}{K}\sum_{k=1}^{K}
      \Bigl(\widehat{q}_{\boldsymbol{\theta}}(\alpha_k)
            - \widehat{p}_{\mathrm{train}}(\alpha_k)\Bigr)^{\!2}.
  \label{eq:loss-parity}
\end{equation}

The coupling this induces is best seen through a single mask. Recall
$\alpha=0110$ from \Cref{sec:background:walsh}: the empirical moment
$\widehat{p}_\mathrm{train}(0110)
  = \tfrac{1}{m}\sum_{x\in\cD_{\mathrm{train}}}\phi_{0110}(x)$
is computed only from observed states, yet matching
$\widehat{q}_\theta(0110)$ to it forces the model to reproduce the
correct agreement tendency of bits~2 and~3 across all $2^n$ states at
once, including those absent from $\cD_{\mathrm{train}}$. Every term in
Equation~\eqref{eq:loss-parity} imposes this kind of global constraint through
a different bit subset, so even a moderate band simultaneously
constrains observed and unobserved regions of the distribution.

As a coordinate-wise counterpart we pair the parity loss with
\begin{equation}
  \mathcal{L}_{\mathrm{MSE}}(\boldsymbol{\theta})
    = \frac{1}{|\cS|}\sum_{x\in\cS}
      \Bigl(q_{\boldsymbol{\theta}}(x)
            - \hat{p}_{\mathrm{train}}(x)\Bigr)^{\!2}.
  \label{eq:loss-mse}
\end{equation}
Both~\eqref{eq:loss-parity} and~\eqref{eq:loss-mse} are $L_2$-type empirical
objectives, but they apply the squared mismatch in different coordinate
systems: \eqref{eq:loss-parity} measures mismatch in Walsh moment space,
whereas \eqref{eq:loss-mse} measures mismatch directly in state space.
The zero target for an unobserved state is not a special weakness of
MSE; it is the standard finite-sample consequence of any unsmoothed
pointwise empirical objective whose per-state targets are sample
frequencies.

For valid states absent from $\cD_{\mathrm{train}}$, the empirical
coordinate-wise target $\hat{p}_{\mathrm{train}}(x)$ is zero because the
state was not sampled. This is the standard finite-sample behavior of
unsmoothed pointwise empirical objectives, not an artificially weak MSE
baseline. The model is not forced to assign exactly zero probability to
such states; rather, the direct per-state term provides no positive
sample evidence for them and penalizes probability assigned there. In
contrast, parity supervision still couples those states to the observed
sample through $\phi_{\alpha_k}(x^\star)$ in every matched moment,
allowing evidence to be transferred whenever their parity fingerprint is
consistent with the observed moment constraints.

Accordingly, our conclusion concerns the unsmoothed coordinate-wise MSE
used here. Smoothed probability targets and non-parity global
moment-matching objectives were not evaluated and could yield a
different comparison.

\subsection{Evaluation Metric}
\label{sec:methodology:evaluation}

Because the benchmark provides access to the full target $p^\star$ by
construction, we evaluate each model by the forward KL divergence from
target to model,
\begin{equation}
  D_{\mathrm{KL}}(p^\star\|q)
    =\sum_{x\in\cS} p^\star(x)\,\log\frac{p^\star(x)}{q(x)},
  \label{eq:kl}
\end{equation}
computed exactly by complete enumeration of $p^\star$ and $q$ over the
valid support $\cS$. Mass assigned outside $\cS$ is only one failure mode. It does not
appear explicitly in \eqref{eq:kl}, but is captured by the
support-leakage term $\log(1/q(\cS))$ in the decomposition of
\Cref{app:kl-decomposition}. The same decomposition also separates
this effect from region-mass misallocation and within-region shape
errors, while \Cref{sec:methodology:coverage} separately evaluates
whether recovered unseen mass lands on the high-value subset $\cE_\tau$
rather than on arbitrary unseen states. We choose forward over reverse KL
because it penalizes the model for placing too little mass on states
the target considers important---the failure mode most relevant to
out-of-sample generalization.

\subsection{High-Value-State Coverage}
\label{sec:methodology:coverage}

In many combinatorial-generation settings, valid states are not equally
useful: beyond fitting the observed data distribution, one often cares
about discovering previously unseen states of high task-specific
quality. We formalize this by equipping each valid state $x$ with a
score function $\ell(x)$ that quantifies its utility within the task.
This raises a question that global fit metrics alone cannot answer:
does a model preferentially place probability mass on useful unseen
states, or does it merely spread mass over unseen states more broadly?

Forward KL captures overall distributional accuracy, but it does not
reveal where a fit improvement localizes. Related combinatorial-generation
work therefore often relies on proxy criteria such as finite-budget
discovery counts~\cite{Nica2022Evaluating,Shen2023GFlowNet,%
Gili2022GeneralizationMetrics,HibatAllah2024PQA}. Because $p^\star$ is
fully known in our controlled benchmark, we can replace such proxies by
an exact set-based discovery calculation tailored to the question of
whether a model places mass on high-value states absent from training,
rather than merely on arbitrary unseen states.

Let $\cO \subseteq \cS$ denote the set of valid states that appear at
least once in $\cD_{\mathrm{train}}$, and let $\cU = \cS \setminus \cO$
be its complement. For a threshold quantile $\tau \in (0,1]$ we define the \emph{high-value region}
\begin{equation}
  \cH_\tau \;=\; \bigl\{\, x \in \cS
    : \ell(x) \ge Q_{1-\tau}(\ell) \bigr\},
  \label{eq:high-value-region}
\end{equation}
where $Q_{1-\tau}(\ell)$ is the $(1{-}\tau)$-quantile of the score
distribution over $\cS$~\cite{Gili2022GeneralizationMetrics,%
HibatAllah2024PQA}. The high-value states absent from training form the
\emph{unseen elite}
\begin{equation}
  \cE_\tau \;=\; \cH_\tau \cap \cU
         \;=\; \cH_\tau \setminus \cO.
  \label{eq:unseen-elite}
\end{equation}
Throughout we report $\tau = 0.1$, i.e.\ the upper-score region defined by the 90th-percentile threshold.

The metric asks how many elements of $\cE_\tau$ a model surfaces under
a finite sampling budget, not how closely it approximates $p^\star$
on those elements; accuracy and discovery are thus disentangled here
and later reconciled with the KL fit in \Cref{sec:results}. If we
draw $Q$ independent samples from $q$, the expected number of distinct
states in $\cE_\tau$ seen at least once is
\begin{equation}
  M_q(Q) \;=\; \sum_{x \in \cE_\tau}
               \Bigl[1 - (1 - q(x))^Q\Bigr],
  \label{eq:expected-discoveries}
\end{equation}
the standard occupancy expression~\cite{Gnedin2007Occupancy}. From it
we derive the two diagnostics used throughout the results,
\begin{equation}
  C_q(Q) \;=\; \frac{M_q(Q)}{Q},
  \qquad
  R_q(Q) \;=\; \frac{M_q(Q)}{|\cE_\tau|},
  \label{eq:coverage-recovery}
\end{equation}
namely discoveries per sample and the cumulative fraction of
$\cE_\tau$ recovered.

\begin{table}[t]
\centering
\caption{Core notation used in the benchmark and diagnostics.}
\label{tab:notation}
\setlength{\tabcolsep}{4pt}
\footnotesize
\begin{tabularx}{\linewidth}{@{}p{0.16\linewidth}X@{}}
\toprule
\textbf{Symbol} & \textbf{Meaning} \\
\midrule
$\cS$        & valid support \\
$\cO$        & observed valid states under $\cD_{\mathrm{train}}$ \\
$\cU$        & unobserved valid states under $\cD_{\mathrm{train}}$ \\
$\tau$       & quantile threshold defining the high-value region \\ 
$\cH_\tau$   & high-value region, states with $\ell(x)\ge Q_{1-\tau}(\ell)$ \\ 
$\cE_\tau$   & unseen high-value states, $\cH_\tau\cap\cU$ \\ 
$\Omega$     & sampled parity band of nonzero Walsh masks \\
$K$          & band width, $K=|\Omega|$, \emph{i.e.}, the number of matched masks/moments \\
$\sigma$     & band-selection parameter controlling the weight profile of $\Omega$ \\
$C_q(Q)$     & discoveries per sample over~$\cE_\tau$ \\
$R_q(Q)$     & cumulative recovery fraction over~$\cE_\tau$ \\ 
$q_{\mathrm{lin}}$  & band-limited linear reconstruction \\
$q_{\mathrm{spec}}$ & clipped and renormalized spectral proxy \\
\bottomrule
\end{tabularx}
\end{table}

\subsection{Spectral Completion and High-Value Region Visibility}
\label{sec:methodology:spectral}

To quantify whether the sampled parity band alone places more than
uniform mass on the unseen elite $\cE_\tau$, we construct a
parameter-free reconstruction from the matched moments and decompose
it into a uniform baseline and a band-induced term. The resulting
\emph{spectral visibility} is the analytic counterpart of the
circuit-free baseline used in \Cref{sec:results} to diagnose how much of
the recovery is already encoded in the parity signal before IQP
parameterization.

Let $\Omega = \{\alpha_1, \dots, \alpha_K\}$ be the sampled set of
nonzero Walsh masks and $\widehat{p}_{\mathrm{train}}(\alpha_k)$ the
corresponding empirical parity moments defined in
Equation~\eqref{eq:parity-moment}. The raw band-limited
reconstruction is
\begin{equation}
  q_{\mathrm{lin}}(x)
    = \frac{1}{2^n}
      \Bigl(1 + \sum_{k=1}^{K}
        \widehat{p}_{\mathrm{train}}(\alpha_k)\,(-1)^{\alpha_k \cdot x}\Bigr),
  \label{eq:q-linear}
\end{equation}
the partial Walsh inversion on the Boolean cube restricted to the
sampled band~\cite{deWolf2008BooleanFourier}. Since $q_{\mathrm{lin}}$
need not be nonnegative, we turn it into a valid probability
distribution by clipping and renormalizing,
\begin{equation}
  q_{\mathrm{spec}}(x)
    = \frac{[q_{\mathrm{lin}}(x)]_+}
           {\sum_{y \in \cS}\,[q_{\mathrm{lin}}(y)]_+},
  \label{eq:q-spectral}
\end{equation}
with $[a]_+ = \max(a, 0)$. We treat $q_{\mathrm{lin}}$ as the analytic
object and $q_{\mathrm{spec}}$ as its valid-probability projection
used in the experiments; both are parameter-free and depend only on
$\Omega$ and $\widehat{p}_{\mathrm{train}}$.

The analytic tractability of $q_{\mathrm{lin}}$ yields a clean
decomposition of the mass it assigns to any region $A \subseteq \cS$.
Defining
\begin{equation}
  \bar\phi_A(\alpha) \;:=\; 2^{-n} \sum_{x \in A} (-1)^{\alpha \cdot x}
  \label{eq:region-character-mean}
\end{equation}
and summing \eqref{eq:q-linear} over $A$ gives the
\emph{uniform-plus-visibility decomposition}
\begin{equation}
  q_{\mathrm{lin}}(A)
    = \underbrace{\frac{|A|}{2^n}}_{\text{uniform baseline}}
    + \underbrace{\sum_{k=1}^{K}
        \widehat{p}_{\mathrm{train}}(\alpha_k)\,\bar\phi_A(\alpha_k)}_{
        \mathrm{Vis}_\Omega(A)}.
  \label{eq:q-linear-region}
\end{equation}

The first term is the mass a uniform distribution would assign to $A$;
the second, which we call the \emph{spectral visibility}
$\mathrm{Vis}_\Omega(A)$, is the additional mass induced by the
matched parity band. A region with $\mathrm{Vis}_\Omega(A) > 0$ is
one the band can already ``see'' as heavier than uniform---hence the
name.

The decomposition in \eqref{eq:q-linear-region} applies to any region
$A \subseteq \cS$. Our primary notion of overall generalization remains
the exact forward-KL fit over the full valid support $\cS$. As a
complementary, task-oriented diagnostic, however, we focus on the
unseen elite $A=\cE_\tau$, because this is the region relevant to the
high-value-state discovery analysis of
\Cref{sec:methodology:coverage}. If $\mathrm{Vis}_\Omega(\cE_\tau) > 0$,
the sampled parity band alone---before any circuit optimization---already
biases mass toward high-value states absent from training. We therefore
use this quantity in \Cref{sec:results} not as a replacement for
global generalization metrics, but as a mechanistic diagnostic showing
that part of the recovery effect is already encoded in the training signal
itself, before any parameters are learned.

\section{Experimental Setup}
\label{sec:setup}

\begin{figure}[t]
\centering
\maybeincludegraphics[width=1\linewidth]{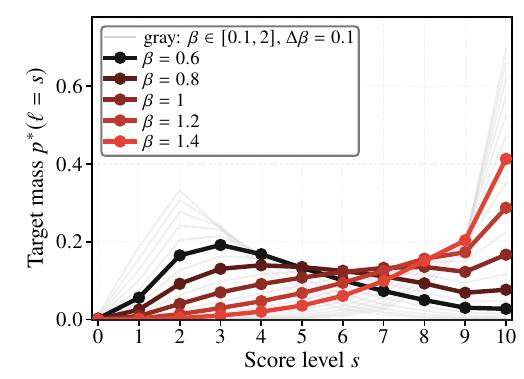}
\caption{Target mass per score level across the $\beta$ sweep for $n=12$. Larger~$\beta$ concentrates mass on fewer high-score states.}
\label{fig:target-sharpness}
\end{figure}

\begin{figure*}[t]
\centering
\maybeincludegraphics[width=\textwidth]{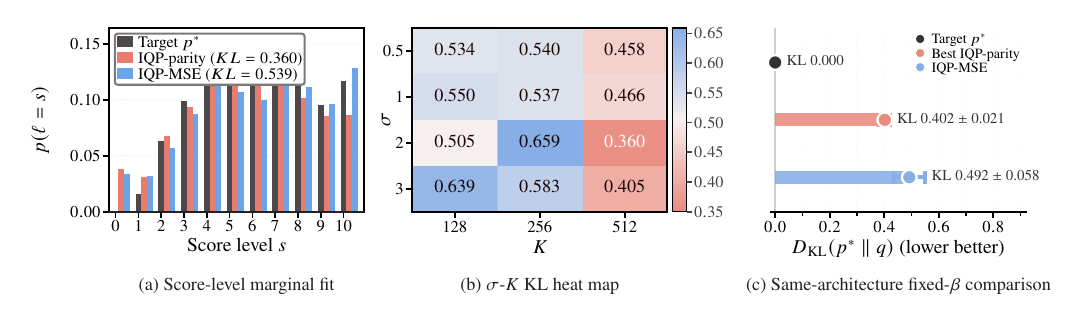}
\vspace{-20pt}
\caption{Exact KL diagnostics at the representative fixed-$\beta$ slice $\beta=0.9$. The integrated figure shows the score-level marginal $p(\ell=s)$ of $p^\star$, the best parity-trained IQP model, and the IQP-MSE control for the representative seed; the exact $D_{\mathrm{KL}}(p^\star\|q)$ values across the 12 parity-band configurations $(\sigma,K)$, with the dashed box marking the best setting; and the parity-trained IQP versus IQP-MSE comparison aggregated over the ten paired instances at $\beta=0.9$, with $p^\star$ as the zero-KL reference.}
\label{fig:overview-kl}
\end{figure*}

We instantiate the benchmark framework of \Cref{sec:methodology} at
$n=12$. The small size is a methodological choice: it allows exact
enumeration of the $2^n$-dimensional output distribution and exact
forward-KL evaluation against the known target. We use this
exponential-time computation only for evaluation and attribution. The
training signal itself---parity moments estimated from circuit samples---does
not require enumerating the state space; each moment is the expectation
of a Pauli-$Z$ string and can be estimated to precision~$\epsilon$ from
$O(1/\epsilon^2)$ samples independently of~$n$ (cf.\ \Cref{sec:background}).
\Cref{tab:benchmark-constants} collects the numerical constants; all
statevector experiments are implemented in JAX.

\subsection{Benchmark Family and Sweep Scope}
\label{sec:setup:benchmark}

The target family lives on the Boolean cube $\{0,1\}^{12}$, with
validity restricted to the even-parity support
\begin{equation}
  \cS
  = \bigl\{x \in \{0,1\}^{12} :
     \textstyle\sum_{i=1}^{12} x_i \equiv 0 \pmod 2 \bigr\},
  \label{eq:valid-support}
\end{equation}
so $|\cS| = 2^{11} = 2048$. We use even parity as a simple but
nontrivial proof-of-concept validity constraint. In many structured
generation tasks, valid data are characterized by global combinatorial
patterns rather than by unstructured frequency fitting alone. Even
parity captures this idea in a minimal and analytically clean form: it
defines a large valid support while still leaving a nontrivial problem
of how probability mass should be distributed within that support. This
makes it a controlled setting for testing whether parity-based
supervision helps a model learn structure beyond mere support
membership, while keeping exact attribution possible.

For the score we use the length of the longest zero block bracketed
by ones, denoted $\ell(x)$, i.e.\ the largest number of consecutive
zeros appearing between two ones. On the even-parity support at
$n=12$, this yields score levels $s\in\{0,\ldots,10\}$. Importantly,
the even-parity constraint defines only the valid support $\cS$; it
does not define the target probabilities within that support, nor is
the score $\ell(x)$ supplied to the loss. The score depends on
variable-length contiguous patterns and is therefore not captured
by any single Walsh coefficient. Instead, its representation is
distributed across multiple Walsh components, so parity-based
training is not given a direct shortcut to the score landscape.

For the distribution family we adopt the Boltzmann form
\begin{equation}
  p^\star_\beta(x)
    = \frac{1}{Z_\beta}\exp\bigl(\beta\,\ell(x)\bigr)\,\ind\{x\in\cS\},
  \label{eq:target-family}
\end{equation}
a canonical choice in combinatorial optimization. The sharpness
parameter~$\beta\in\{0.1,0.2,\ldots,2.0\}$ (20 values, step size
$\Delta\beta=0.1$) acts as a single difficulty control: small~$\beta$
yields a near-uniform target that is easy to approximate, while
large~$\beta$ concentrates mass on a shrinking set of high-score
states (\Cref{fig:target-sharpness}). With a training-sample
size of $m=200$ per instance, the benchmark operates in a
data-scarce regime: each training set observes less than $10\%$
of~$\cS$, so generalization to unseen valid states is nontrivial
by construction.

\begin{table}[t]
\centering
\caption{Benchmark, training, and evaluation constants used throughout the paper.}
\label{tab:benchmark-constants}
\small
\begin{tabularx}{\linewidth}{@{}p{0.35\linewidth}X@{}}
\toprule
\textbf{Quantity} & \textbf{Value} \\
\midrule
\multicolumn{2}{@{}l}{\emph{Benchmark family}}\\
System size            & $n=12$, even-parity support, $|\cS|=2048$ \\
Sharpness sweep        & $\beta\in\{0.1,0.2,\ldots,2.0\}$ \\
Score function         & longest zero block bracketed by ones, $\ell(x)$ \\ 
\midrule
\multicolumn{2}{@{}l}{\emph{Training}}\\
Training-sample size   & $m=200$ independent samples from $p^\star_\beta$ \\
Paired instances       & $200$ ($20\,\beta \times 10$ seeds) \\
Reference parity band  & $(\sigma,K)=(1,512)$ \\
Fixed-$\beta$ ablation & $\beta=0.9$;\; $\sigma\in\{0.5,1,2,3\}$,\; $K\in\{128,256,512\}$ \\
\midrule
\multicolumn{2}{@{}l}{\emph{Evaluation}}\\
High-value region      & $\tau=0.1$, i.e.\ $\ell(x)\ge Q_{0.9}(\ell)$ \\
Coverage budgets       & $Q\in\{1000,2000,5000\}$ \\
\bottomrule
\end{tabularx}
\end{table}

\subsection{Paired-Instance Protocol}
\label{sec:setup:protocol}

The unit of comparison is a \emph{paired instance} $i=(\beta,r)$
with $\beta\in\{0.1,\ldots,2.0\}$ and ten seed IDs $111,\ldots,120$, for a
total of $200$ instances. Within an instance, all compared models
share the same training multiset~$\cD_{\mathrm{train}}$ (drawn independently
from~$p^\star_\beta$) and, for parity-based methods, the same
Walsh-mask band~$\Omega$; only the model initialization differs.

The main sweep uses the reference band $(\sigma,K)=(1,512)$
across all $200$ instances. A dedicated fixed-$\beta$ ablation at
$\beta=0.9$ varies $\sigma\in\{0.5,1,2,3\}$ and
$K\in\{128,256,512\}$, and \Cref{app:kl-coverage-n-sweep}
extends the fixed-$\beta$ ordering to $n\in\{10,\ldots,20\}$ under
the same paired-instance protocol.

\subsection{Models, Training, and Metric Computation}
\label{sec:setup:models}

\paragraph{Controlled comparison design}
The within-IQP contrast holds architecture, data, optimizer, and update
budget fixed while swapping parity supervision for coordinate-wise MSE,
which isolates the loss effect. The cross-model contrast compares
IQP-parity with the maximum-entropy parity model (MaxEnt-parity) and sparse
Ising under matched parity information. MaxEnt-parity receives the same Walsh moments as IQP-parity
and no additional information. Further baselines provide external
reference points. Together, these comparisons provide diagnostic
evidence about model-class effects.

\paragraph{Classical baselines}
As classical baselines, we use two \emph{classical pairwise Ising
models} and one \emph{autoregressive (AR) Transformer}. The sparse NN{+}NNN
Ising model with local fields matches the interaction pattern of the IQP
circuit and is trained on the same sampled parity-moment objective as
IQP-parity. The dense Ising model includes all pairwise couplings and
local fields and is trained by empirical cross entropy. The
autoregressive Transformer is trained by maximum likelihood and provides
a non-energy-based likelihood baseline. These controls cover
parity-aware pairwise fitting, dense pointwise energy modeling, and
autoregressive likelihood modeling. Each model is sized independently.
Several Transformer capacities were evaluated in preliminary
comparisons. The medium configuration performed best overall and was
fixed for all main sweeps. Full configurations are listed in
\Cref{tab:model-configurations}; training details are given in
\Cref{app:classical_baseline_training}. We do not claim that these
baselines exhaust all possible classical parity-aware models.

\paragraph{Training and fairness}
Fairness is enforced through shared data and matched training budgets
rather than equal parameter counts. IQP-parity and IQP-MSE use the same
training data, circuit, optimizer, and 600 update steps, so their
comparison isolates the loss. Optimization settings for all models are
reported in \Cref{app:classical_baseline_training}.

\paragraph{Metric computation}
We exactly enumerate every trained model over all $2^n$ bit strings,
yielding complete probability tables for forward~KL, coverage, and
recovery. The high-value region is the $\tau=0.1$ upper-score region of
$\cS$ under $\ell(x)$; the unseen elite~$\cE_\tau$
\eqref{eq:unseen-elite} is the subset absent from the training sample.
The band-limited reconstruction~$q_{\mathrm{spec}}$ is evaluated in the
same way but serves as a parameter-free diagnostic for the parity signal
(\Cref{sec:methodology:spectral}), not as a trained competitor.

\begin{table}[t]
\centering
\caption{Model configurations and capacity summary. IQP-parity and
IQP-MSE share the same circuit and differ only in supervision.}
\label{tab:model-configurations}
\setlength{\tabcolsep}{3pt}
\footnotesize
\begin{tabularx}{\linewidth}{@{}m{0.17\linewidth}m{0.26\linewidth}X>{\centering\arraybackslash}m{0.17\linewidth}@{}}
\toprule
\textbf{Model} & \textbf{Architecture} & \textbf{Objective} & \textbf{Parameters} \\
\midrule
IQP-parity     & IQP ring, NN{+}NNN couplings    & parity loss, \eqref{eq:loss-parity} & 24 \\[4pt]
IQP-MSE        & identical circuit               & MSE, \eqref{eq:loss-mse}             & 24 \\[4pt]
MaxEnt-parity  & sampled parity moments          & maximum entropy (convex)              & 512$^{\mathrm{a}}$ \\[4pt]
Sparse Ising+fields & NN{+}NNN graph, fields          & parity loss, \eqref{eq:loss-parity} & 36 \\[4pt]
Dense Ising+fields & dense graph, fields          & cross-entropy                         & 78 \\[4pt]
AR Transformer & autoregressive (medium)$^{\mathrm{b}}$ & maximum likelihood               & 9{,}057 \\[4pt]
$q_{\mathrm{spec}}$ & band-limited reconstruction & non-iterative diagnostic            & 0 \\
\bottomrule
\end{tabularx}
\par\vspace{7pt}
\begin{minipage}{\linewidth}
\footnotesize
$^{\mathrm{a}}$~Scales with $K$; 512 at the reference band width $K\!=\!512$.\\
$^{\mathrm{b}}$~$d_{\mathrm{model}}\!=\!32$, one layer, four heads, $d_{\mathrm{ff}}\!=\!64$. Fixed main-sweep configuration after preliminary capacity comparisons.
\end{minipage}
\end{table}

\section{Results}
\label{sec:results}

\begin{figure*}[!ht]
\centering
\maybeincludegraphics[width=\textwidth]{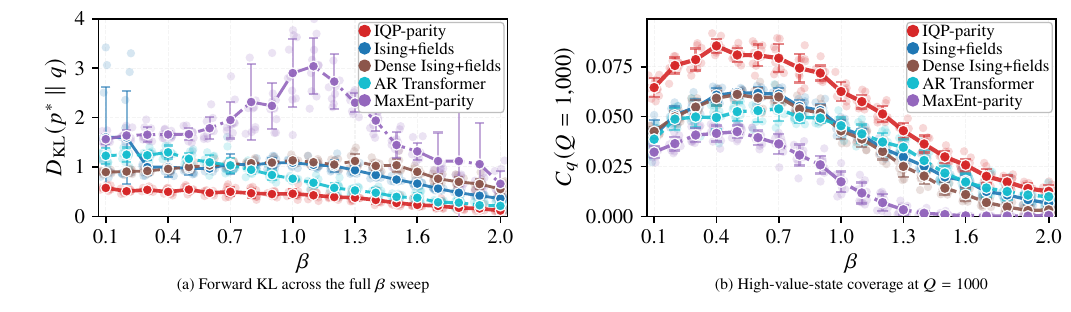}
\caption{Cross-class diagnostics under the reference parity band $(\sigma,K)=(1,512)$. (a) Lower KL is better; (b) higher coverage is better.}
\label{fig:beta-sweep-coverage}
\end{figure*}

The results are organized as an attribution argument. We first compare two losses inside the same IQP circuit (\Cref{sec:results:loss}), then compare the resulting IQP-parity model with the tested classical controls (\Cref{sec:results:crossed-design}), then ask whether the fit improvement translates into high-value-state discovery (\Cref{sec:results:coverage}), and finally use the parameter-free spectral proxy to identify how much of the recovery originates in the training signal itself (\Cref{sec:results:spectral}).

\subsection{Inside the Same IQP Circuit, Parity Supervision Improves over Coordinate-Wise Fitting}
\label{sec:results:loss}

\paragraph{Same circuit, different loss}
Holding architecture, data, and optimization budget fixed, replacing the MSE loss with parity supervision reduces mean forward KL from $0.49\!\pm\!0.06$ to $0.40\!\pm\!0.02$ at $\beta{=}0.9$, an 18\% improvement. In the broader cross-class comparison of \Cref{tab:full_sweep_raw_summary}, IQP-parity attains the lowest KL in 190 of 200 paired instances.

This is the cleanest isolation of the training signal in our design: the two IQP models share the same 24-parameter circuit, the same training data, the same optimizer, and the same 600-step budget. Only the loss differs. The observed gap is therefore attributable to the supervision signal rather than to circuit capacity or data access.

\Cref{fig:overview-kl} makes the within-configuration picture concrete. At the representative slice $\beta=0.9$, the central heat map encodes exact $D_{\mathrm{KL}}(p^\star \| q)$, with lower values indicating better fit. All four $K=512$ parity bands and several of the $K=128$ and $K=256$ bands outperform the IQP-MSE control (KL $=0.54$). The best single configuration reaches $0.36$, a 33\% reduction relative to MSE. The three-panel figure also shows the score-level marginal fit and the seed-wise comparison at this $\beta$, confirming that the effect is not a seed artifact.

\paragraph{Why the loss matters}
Both objectives see the same training histogram, but they turn it into different constraints. For an unobserved state, the unsmoothed coordinate-wise target in \eqref{eq:loss-mse} is zero simply because the state was not sampled; the direct term therefore provides no positive per-state sample evidence and penalizes probability assigned there. This does not force the model to assign exactly zero probability. In \eqref{eq:loss-parity}, each matched Walsh coefficient is global: it couples all $2^n$ states through $\phi_\alpha$, so an unobserved state whose parity fingerprint aligns with observed structure can receive evidence through the shared moment constraints. This mechanism reappears below as the spectral-completion diagnostic.

\subsection{IQP-Parity Outperforms the Tested Baselines in the Controlled Setting}
\label{sec:results:crossed-design}

\paragraph{Ranking across the full sweep}
Under the reference band $(\sigma,K)=(1,512)$, IQP-parity attains the lowest forward KL at every one of the 20 tested $\beta$ values, as shown in the left panel of \Cref{fig:beta-sweep-coverage}. The tested classical baselines---two Ising variants, a tuned autoregressive Transformer, and a MaxEnt model trained on the identical parity moments---yield higher forward-KL values across the sweep under this metric.

\begin{table}[t]
\centering
\caption{
Summary across 20 $\beta$ values $\times$ 10 seeds (200 paired instances).
Mean $\pm$ 95\% confidence interval (CI); medians report the skewed KL distribution. ``KL wins''
counts instances on which the model attains the lowest KL among the five classes.
}
\label{tab:full_sweep_raw_summary}
\setlength{\tabcolsep}{4pt}
\footnotesize
\begin{tabularx}{\linewidth}{@{}>{\raggedright\arraybackslash}Xcccc@{}}
\toprule
\textbf{Model} & \textbf{mean KL} & \textbf{median KL} & \textbf{KL wins} & \textbf{mean $C_q(1000)$} \\
\midrule
\textbf{IQP-parity} & $\mathbf{0.385 \pm 0.021}$ & $\mathbf{0.414}$ & $\mathbf{190/200}$ & $\mathbf{0.053 \pm 0.004}$ \\
Ising+fields (NN\allowbreak+\allowbreak NNN) & $0.923 \pm 0.062$ & $0.929$ & $0/200$ & $0.038 \pm 0.003$ \\
Dense Ising+fields & $0.947 \pm 0.025$ & $0.978$ & $0/200$ & $0.035 \pm 0.003$ \\
AR Transformer & $0.744 \pm 0.054$ & $0.737$ & $10/200$ & $0.036 \pm 0.002$ \\
MaxEnt-parity & $1.804 \pm 0.108$ & $1.689$ & $0/200$ & $0.018 \pm 0.002$ \\
\bottomrule
\end{tabularx}
\end{table}

\paragraph{What the controlled comparisons tell us}
The within-IQP loss swap isolates the supervision effect for the fixed
IQP circuit, while the cross-model comparisons provide diagnostic
evidence about model class. IQP-MSE shows that the circuit alone does
not reproduce the parity-supervised result, while MaxEnt-parity shows
that parity moments alone are insufficient in the canonical
least-biased classical control. The spectral proxy attributes much of
the recovery signal to parity supervision, while the IQP
parameterization refines it into the strongest fit among the tested
models.

\paragraph{Robustness}
\Cref{app:kl-coverage-n-sweep} repeats the comparison in a size sweep from $n=10$ to $n=20$ at fixed $\beta$. Under the same protocol and without per-$n$ retuning, IQP-parity yields the lowest median KL at each tested system size. Because the KL distribution is right-skewed at small $\beta$, we report both the mean with 95\% CI and the median; the qualitative ordering is unchanged under either summary.

\subsection{The Fit Improvement Translates to Better Unseen High-Value-State Discovery}
\label{sec:results:coverage}

\begin{figure*}[!ht]
\centering
\maybeincludegraphics[width=\textwidth]{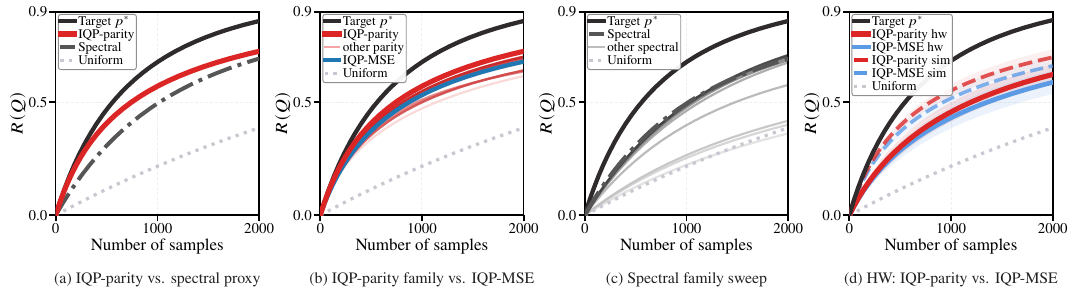}
\vspace{-20pt}
\caption{Mechanism at $\beta{=}0.9$. Cumulative recovery $R_q(Q)$ on the unseen high-value set $\cE_\tau$. (a) The parameter-free spectral proxy $q_{\mathrm{spec}}$ already captures much of the recovery gain of the best IQP-parity model over uniform sampling. (b) Within the IQP family, parity-trained variants recover $\cE_\tau$ faster than IQP-MSE. (c) The parameter-free proxy exhibits the same $(\sigma,K)$ spread. This is consistent with a strong influence of the selected band, rather than circuit optimization alone, on the ordering. (d) The loss-level ordering remains visible on IBM hardware (HW), with bands showing $\pm 1$ standard deviation over 10 matched seeds.}
\label{fig:spectral-mechanism}
\end{figure*}

\paragraph{From global fit to operational discovery}
Lower KL translates into discovering more unseen high-value states in this setting. At $Q{=}1000$ samples, IQP-parity achieves $C_q(1000)=0.053 \pm 0.004$, i.e.\ about $53$ distinct unseen high-value states per $1000$ samples on average, versus about $35$--$38$ for the trained classical baselines and $18$ for MaxEnt-parity (\Cref{tab:full_sweep_raw_summary}, rightmost column).

A global score could in principle improve in irrelevant regions. The right panel of \Cref{fig:beta-sweep-coverage} shows that here the KL ranking and the coverage ranking coincide across the $\beta$ sweep. 

\paragraph{The shape of the $\beta$ curve is informative}
Coverage peaks at intermediate sharpness. When $\beta$ is small the target is near-uniform and the high-value region is only weakly distinguished, so all models perform similarly. When $\beta$ is large the target concentrates on a handful of heavy states, and repeated hits on those few states suppress the rate at which additional unseen states are uncovered. Between these regimes, where the structural problem is hardest, the separation among methods is widest.

\subsection{Spectral Completion Shows That Much of the Recovery Comes From the Training Signal}
\label{sec:results:spectral}

\paragraph{The parity band already generalizes before circuit training}
A parameter-free reconstruction $q_{\mathrm{spec}}$ built from the sampled band moments alone (\Cref{sec:methodology:spectral}) already recovers much of the gap between IQP-parity and uniform sampling, as shown in the leftmost panel of \Cref{fig:spectral-mechanism}. The IQP circuit then refines this scaffold.

This diagnostic shows where much of the recovery originates: parity moments computed from the training sample encode a completion rule, transferring evidence to unobserved states whose parity fingerprints match observed structure. This is the mechanism from \Cref{sec:results:loss}, now visible without circuit optimization.

Two further panels support this view. The second panel shows that within the IQP family, parity-trained variants recover $\cE_\tau$ faster than IQP-MSE. The third panel shows that $q_{\mathrm{spec}}$ exhibits the same qualitative spread across $(\sigma,K)$ choices as the trained models, indicating that the band itself strongly shapes the ordering.

\paragraph{The loss-level ordering remains visible on hardware}
The rightmost panel of \Cref{fig:spectral-mechanism} is the small-scale IBM hardware feasibility check described in \Cref{app:reproducibility:hardware}. Averaged over 10 matched seeds, the parity-trained IQP model retains the same loss-level ordering over IQP-MSE despite shot noise and device imperfections; the shaded bands ($\pm 1$ standard deviation) do not overlap over most of the budget range. The hardware curves sit below the simulator curves, as expected. We do not interpret this as a hardware-scale advantage claim, but as evidence that the effect is not limited to noiseless statevector simulation.

\paragraph{How the circuit complements the spectral scaffold}
The leftmost panel of \Cref{fig:spectral-mechanism} shows a remaining
gap between IQP-parity and $q_{\mathrm{spec}}$. Since
$q_{\mathrm{spec}}$ is only a parameter-free diagnostic of the sampled
parity band, this gap suggests that the IQP ansatz adds structure
beyond the band-limited reconstruction, for example through interactions
among Walsh coefficients outside~$\Omega$. \Cref{app:kl-decomposition}
provides the KL decomposition used to interpret the final errors in terms
of support leakage, observed--unobserved mass allocation, and conditional
shape errors.

\medskip
\noindent Together, the four subsections support the following conclusion: in this controlled parity-structured setting, parity supervision provides a spectral completion signal, and the IQP parameterization refines that signal into the best distributional fit among the tested models.

\section{Discussion and Scope}
\label{sec:discussion}

The broader implication is that the parity structure used in IQP training should not be viewed only as a technical restriction. When the data distribution has compatible spectral structure, the same parity signal that makes IQP-MMD-style training tractable can also couple observed and unobserved states through shared Walsh fingerprints. In this sense, parity supervision is both a computationally accessible training signal and, in the tested regime, a mechanism for generalization beyond the finite training histogram.

Real data could exhibit useful alignment when global constraints or
recurring subset correlations concentrate in low- or moderate-weight
Walsh coefficients. If the structure is diffuse or the selected masks
miss it, parity supervision need not help. Identifying informative masks
from domain knowledge or validation data remains open.

The experiments do not establish an advantage for targets without
compatible Walsh structure, and no non-parity benchmark was evaluated.
Smoothed coordinate-wise targets and non-parity global moments could
also alter the comparison and remain untested.

These findings apply to a deliberately controlled setting: a small,
synthetic, exactly enumerable, parity-aligned benchmark and
non-exhaustive classical baselines. We therefore do not claim practical
quantum advantage or general-purpose superiority of IQP Born machines.
Rather, the contribution is a clean isolation of the loss effect within
the fixed IQP architecture, complemented by diagnostic cross-model
comparisons under matched data and training budgets.

At larger sizes, parity moments remain sample-estimable, but mask
selection, optimization, and evaluation no longer scale as in this
enumerable benchmark. Exact forward KL and recovery must then be
replaced by sample-based or problem-specific estimators.

\section{Conclusion and Future Work}
\label{sec:conclusion}

This work studied whether the parity structure used in IQP Born machines is merely a device for tractable training, or whether it can also act as an inductive bias for generalization. In a controlled, exactly enumerable setting, we find evidence for the latter. Parity supervision improves exact forward-KL fit and unseen high-value-state recovery over a coordinate-wise empirical objective within the same IQP circuit. A parameter-free spectral reconstruction shows that much of this recovery is already encoded in the sampled parity moments, while the IQP parameterization further refines this spectral scaffold. The hardware experiment preserves the same loss-level ordering, indicating that the effect is not confined to noiseless statevector simulation.

Future work should test how far this mechanism extends beyond the controlled regime considered here. Natural directions include adaptive selection of parity bands, richer validity constraints, larger systems where exact evaluation must be replaced by statistical estimators, and application domains with naturally occurring parity-like or spectral structure. If the effect persists in such settings, parity supervision would be more than a convenient route to tractable IQP training: it would provide a principled design pattern for quantum generative models with spectrally aligned objectives and target structure.

\appendices
\crefalias{section}{appendix}
\crefalias{subsection}{appendix}

\section{KL Decomposition and Robustness}
\label{app:kl-decomposition}
\label{app:kl-coverage-n-sweep}

\begin{figure}[!t]
\centering
\maybeincludegraphics[width=\linewidth]{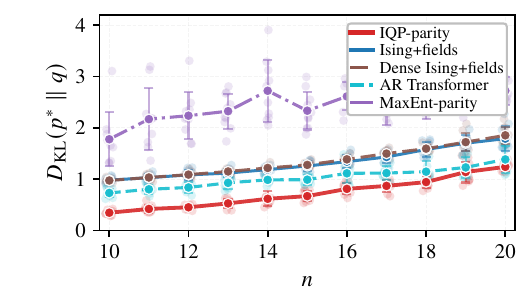}
\caption{Exact forward KL at fixed $\beta=0.9$ across
$n\in\{10,\ldots,20\}$, shown as paired-instance point clouds for all
baselines. IQP-parity achieves the lowest median KL at every tested size
without per-$n$ retuning.}
\label{fig:n-sweep-all-baselines}
\end{figure}

Let $a=p^\star(\cU)$ with $\cU=\cS\setminus\cO$ and
$b=q(\cU)/q(\cS)$. The conditional-KL split used in the diagnostics is
\begin{equation}
\label{eq:kl-decomposition}
\begin{aligned}
D_{\mathrm{KL}}(p^\star\|q)
&=\log \frac{1}{q(\cS)}+D_{\mathrm{KL}}(\mathrm{Bern}(a)\|\mathrm{Bern}(b))\\
&\quad+aD_{\mathrm{KL}}(p^\star_{\cU}\|q_{\cU})+(1-a)D_{\mathrm{KL}}(p^\star_{\cO}\|q_{\cO}).
\end{aligned}
\end{equation}
The four terms in \eqref{eq:kl-decomposition} separate support leakage, observed/unobserved mass
splitting, unseen-region shape error, and observed-region shape error.
Across the full $\beta$ sweep, lower exact forward KL aligns with
higher unseen high-value-state coverage. At fixed $\beta=0.9$, a size
sweep over $n\in\{10,\ldots,20\}$ with ten paired instances per size
and no per-$n$ retuning preserves the ordering: IQP-parity has the
lowest median KL at every tested $n$, rising from $0.34$ at $n=10$ to
$1.26$ at $n=20$. \Cref{fig:n-sweep-all-baselines} shows the paired-instance
point cloud for all baselines, and \Cref{tab:appendix_n_sweep_compact}
reports representative median KL values for selected system sizes.

\begin{table}[!b]
\centering
\caption{Representative median forward KL $D_{\mathrm{KL}}(p^\star\Vert q)$
at $\beta=0.9$ across 10 matched seeds; lower is better.}
\label{tab:appendix_n_sweep_compact}
\begin{tabular}{lccc}
\toprule
Model & $n=10$ & $n=15$ & $n=20$ \\
\midrule
IQP-parity & \textbf{0.338} & \textbf{0.638} & \textbf{1.256} \\
AR Transformer & 0.720 & 0.957 & 1.318 \\
Ising+fields & 0.981 & 1.266 & 1.670 \\
Dense Ising+fields & 0.970 & 1.266 & 1.828 \\
MaxEnt-parity & 1.707 & 2.251 & 2.633 \\
\bottomrule
\end{tabular}
\end{table}

\section{Classical Baseline Training Details}
\label{app:classical_baseline_training}

All classical baselines follow the paired protocol and reference band in
\Cref{sec:setup:protocol}. Each model is trained once from its prescribed
initialization, without restarts, early stopping, validation selection,
or post-hoc tuning. Masks are sampled from $\pi_\sigma$, with $0^n$
resampled. We define $P_{k,x}=(-1)^{\alpha_k\cdot x}$ and
$\hat z=P\hat p_{\mathrm{train}}$.

Sparse Ising minimizes $K^{-1}\|\hat z-Pq_\theta\|_2^2$ with PennyLane
Adam ($\eta=0.05$, 600 steps). Dense Ising minimizes empirical cross
entropy with the same optimizer and budget and no regularization.
MaxEnt-parity, $q_\theta(x)\propto\exp(\theta^\top P_{\cdot,x})$, is
initialized at $\theta=0$ and trained on its log-partition objective with
PyTorch Adam ($\eta=0.05$, 600 steps).

The Transformer factorizes $q(x)=\prod_i q(x_i\mid x_{<i})$ in big-endian
order and uses the architecture in \Cref{tab:model-configurations} with
vocabulary $\{0,1,\mathrm{BOS}\}$ (BOS denotes beginning of sequence),
Gaussian error linear unit (GELU) activations, zero dropout, and no weight decay.
It is trained by next-bit binary cross entropy with PyTorch Adam
($\eta=10^{-3}$), batch size 256, and 600 full-batch epochs ($m=200$).
All models are evaluated exactly as described in \Cref{sec:setup:models}.

\section{Hardware Execution}
\label{app:reproducibility:hardware}

Twenty 10,000-shot jobs ran on \texttt{ibm\_marrakesh} with Qiskit 2.3.0
and IBM Runtime 0.45.1. Optimization level 1 yielded depths 250--275
and 90--117 two-qubit gates.

\clearpage
\balance
\section*{Acknowledgment}
This work was supported by the LMU Sustainability Fund (EfOiE), the German Federal Ministry of Research, Technology and Space (BMFTR) through QuCUN, QuaRDS, CAQAO, and the Quantum Ecosystems funding program under reference 13N17386, the Munich Quantum Valley consortia K5 and K7, and the Bavarian Ministry of Economic Affairs project 6GQT.
\ifdefined\CPSBuild
  \bibliographystyle{IEEEtran}
\else
  \bibliographystyle{IEEEtranArxiv}
\fi
\bibliography{bstcontrol,references}

\end{document}